\documentclass[prd,aps,showpacs,nofootinbib,eqsecnum,onecolumn,groupedaddress]{revtex4}
\usepackage{amsmath} 
\usepackage{amssymb}
\usepackage{amsfonts}
\usepackage{graphicx,bm}
\usepackage{dcolumn}
\usepackage{color,amsxtra}
\usepackage{epsf}
\usepackage{enumerate}
\usepackage{hhline}
\usepackage{array}
\usepackage{tabularx}
\usepackage{subfigure}

\usepackage{fancyhdr}
\usepackage{mathrsfs}

\newcommand{\be}{\begin{equation}}
\newcommand{\ee}{\end{equation}}
\newcommand{\bea}{\begin{eqnarray}}
\newcommand{\eea}{\end{eqnarray}}
\newcommand{\beaa}{\begin{eqnarray*}}
\newcommand{\eeaa}{\end{eqnarray*}}

\allowdisplaybreaks[4]

\newcommand{\Eqn}[1]{&\hspace{-0.2em}#1\hspace{-0.2em}&}

\def\be{\begin{equation}}
\def\ee{\end{equation}}
\def\bea{\begin{eqnarray}}
\def\eea{\end{eqnarray}}

\begin{document}

\title{Cosmological Issues in $F(T)$ Gravity Theory} 

\author{Kazuharu Bamba$^{1, 2, 3}$}
\affiliation{
$^1$Leading Graduate School Promotion Center,
Ochanomizu University, Tokyo 112-8610, Japan\\
$^2$Department of Physics, Graduate School of Humanities and Sciences, Ochanomizu University, Tokyo 112-8610, Japan\\ 
$^3$Division of Human Support System, Faculty of Symbiotic Systems Science, Fukushima University, Fukushima 960-1296, Japan
}


\begin{abstract} 
We review recent developments on cosmology in extended teleparallel gravity, called ``$F(T)$ gravity'' with $T$ the torsion scalar in teleparallelism. 
We explore various cosmological aspects of $F(T)$ gravity 
including the evolution of the equation of state for the universe, 
finite-time future singularities, thermodynamics, and four-dimensional effective $F(T)$ gravity theories coming from the higher-dimensional Kaluza-Klein (KK) and Randall-Sundrum (RS) theories. 
\end{abstract}

\hspace{13.0cm} OCHA-PP-333, FU-PCG-1

\maketitle

\def\thesection{\Roman{section}}
\def\theequation{\Roman{section}.\arabic{equation}}

\section{Introduction}

The fact that the current accelerated 
the cosmic expansion is currently accelerating has 
been supported by various recent cosmological observations 
including Type Ia Supernovae (SNe Ia), 
cosmic microwave background (CMB) radiation, 
baryon acoustic oscillations (BAO), 
large scale structure (LSS), 
and weak lensing effects (see recent results acquired from 
the Planck satellite~\cite{Planck:2015xua,Ade:2015rim} as well as 
the Wilkinson Microwave anisotropy probe (WMAP)~\cite{Komatsu:2010fb,Hinshaw:2012aka}). 
There exist the following two main procedures to account for the 
late-time cosmic acceleration: 
The introducttion of ``dark energy'' and 
the extension of gravity, e.g., the so-called $F(R)$ gravity 
(for recent reviews on dark energy and modified gravity, 
see, for example,~\cite{Bamba:2012cp} and~\cite{Nojiri:2010wj,Nojiri:2006ri,Capozziello:2010zz,Capozziello:2011et,Bamba:2012yr,delaCruzDombriz:2012xy,Bamba:2013iga,Bamba:2014eea,Joyce:2014kja,Bamba:2015uma}, respectively). 

As a formulation for gravity, 
there has been proposed ``teleparallelism'' where the gravity theory 
is described by using the Weitzenb\"{o}ck 
connection (for a recent detailed review, see~\cite{Aldrovandi:2013wha}). 
This has been considered to be an alternative gravitational theory 
to general relativity. 
This gravity theory is written with the torsion scalar $T$, and 
not the scalar curvature $R$ defined with the Levi-Civita 
connection~\cite{Hehl:1976kj,Hayashi:1979qx,Flanagan:2007dc} as in general relativity. 
Recently, it has been found that as in $F(R)$ gravity, 
not only inflation in the early universe~\cite{Ferraro:2006jd,Ferraro:2008ey} but also the late-time cosmic acceleration~\cite{Bengochea:2008gz,Linder:2010py,Bamba:2010wb,Bamba:2010iw} can occur 
in the so-called $F(T)$ gravity, which is an extended version of 
the original teleparallelism. 

In this paper, we review main cosmological consequences 
in $F(T)$ gravity obtained in Refs.~\cite{Bamba:2010wb,Bamba:2010iw,Bamba:2012vg,Bamba:2013fta}. 
First, we investigate the evolution of the equation of state (EoS) for dark energy~\cite{Bamba:2010wb,Bamba:2010iw}. 
We construct an $F(T)$ gravity model in which 
the crossing of the phantom divide can happen\footnote{In Refs.~\cite{Bamba:2008hq,Bamba:2010bm,Bamba:2009kc,Bamba:2009ay,Bamba:2009vq,Bamba:2009dk,Bamba:2010iy,Bamba:2010zz}, 
such an $F(R)$ gravity model with the crossing of the phantom divide has been 
reconstructed.}. This phenomenon has been suggested with cosmological 
observations in Refs.~\cite{Alam:2004jy,Nesseris:2006er,Wu:2006bb,Alam:2006kj,Jassal:2006gf}. 
Second, we demonstrate that 
the finite-time future singularities~\cite{Nojiri:2005sx,Bamba:2008ut,Bamba:2009uf} can appear in $F(T)$ gravity~\cite{Bamba:2012vg}. In addition, $F(T)$ gravity models with realizing the finite-time future singularities are reconstructed. 
We find that the finite-time future singularities can be cured by 
adding a power-law term $T^\beta$ with $\beta>1$, for instance, 
a $T^2$ term. The same approach has been used for Loop quantum cosmology~\cite{Bamba:2012ka}. 
Furthermore, we examine $F(T)$ models in which 
inflation, the $\Lambda$CDM model, Little Rip scenario~\cite{Frampton:2011sp,Brevik:2011mm,Frampton:2011rh,Astashenok:2012tv,Granda:2011kx,Ivanov:2011np,Ito:2011ae,Belkacemi:2011zk,Xi:2011uz,Makarenko:2012gm}, and 
Pseudo-Rip scenario~\cite{Frampton:2011aa} can be realized. 
Third, we derive four-dimensional effective $F(T)$ gravity theories 
from the five-dimensional Kaluza-Klein (KK)~\cite{F-M,MKK-A-C-F,Overduin:1998pn} and Randall-Sundrum (RS)~\cite{Randall:1999ee,Randall:1999vf} models~\cite{Bamba:2013fta}. It is also demonstrated that inflation and the late-time cosmic 
acceleration can occur in the former four-dimensional effective $F(T)$ gravity 
theory and the latter RS model, respectively. 
We use units of $k_\mathrm{B} = c = \hbar = 1$ and denote the
gravitational constant $8 \pi G$ by 
${\kappa}^2 \equiv 8\pi/{M_{\mathrm{Pl}}}^2$ 
with the Planck mass of $M_{\mathrm{Pl}} = G^{-1/2} = 1.2 \times 10^{19}$GeV.

The paper is organized as follows. 
In Sec.\ II, we consider the cosmological evolution of the 
EoS for dark energy. 
In Sec.\ III, 
we analyze the finite-time future singularities 
and reconstruct $F(T)$ gravity models where these finite-time future singularities can appear. We also examine $F(T)$ models to realize various 
cosmological scenarios. 
In Sec.\ IV, we deduce four-dimensional effective $F(T)$ gravity theories 
from both the KK theories and the RS models. 
In Sec.\ V, summary is presented. 

\section{Cosmological evolutions}

In this section, we explain $F(T)$ gravity and examine 
the cosmological evolutioin of the EoS for dark energy 
based on the main results in Ref.~\cite{Bamba:2010wb}. 
The purpose is to construct an $F(T)$ gravity model in 
which the crossing of the phantom divide can occur 
as suggested by resent cosmologcal observations. 

\subsection{Teleparallelism}

We first explain the formulation of teleparallelism. 
The metric is described as 
$
g_{\mu\nu}=\eta_{A B} e^A_\mu e^B_\nu 
$. 
Here, $\eta_{A B}$ is the metric in the Minkowski space-time, 
$e_A (x^{\mu})$ are orthonormal tetrad components 
($A = 0, 1, 2, 3$) at points $x^{\mu}$ of the manifold 
in the tangent space, 
$\mu, \, \nu = 0, 1, 2, 3$ show coordinate indices on the manifold, 
and$e_A^\mu$ corresponds to the tangent vector of the manifold. 
The Lagrangian is written with the torsion scalar $T$. 
This is different from the case for 
general relativity, in which the Lagrangian is expressed by 
using the scalar curvature $R$. 
The torsion scalar $T$ is defined as 
$T \equiv S_\rho^{\verb| |\mu\nu} T^\rho_{\verb| |\mu\nu}$, 
where 
$T^\rho_{\verb| |\mu\nu} \equiv e^\rho_A 
\left( \partial_\mu e^A_\nu - \partial_\nu e^A_\mu \right)$ 
is the torsion tensor 
and 
$S_\rho^{\verb| |\mu\nu} \equiv \left(1/2\right) 
\left(K^{\mu\nu}_{\verb|  |\rho}+\delta^\mu_\rho \ 
T^{\alpha \nu}_{\verb|  |\alpha}-\delta^\nu_\rho \ 
T^{\alpha \mu}_{\verb|  |\alpha}\right)
$ with
$K^{\mu\nu}_{\verb|  |\rho} \equiv -\left(1/2\right) 
\left(T^{\mu\nu}_{\verb|  |\rho} - T^{\nu \mu}_{\verb|  |\rho} - 
T_\rho^{\verb| |\mu\nu}\right)$ the contorsion tensor. 

The Lagrangian of pure teleparallelism is 
written by the torsion scalar $T$. 
This has been extended to an appropriate function of 
$T$ to realize inflation and the late-time cosmic acceleration. 
This concept is the same as $F(R)$ gravity, 
where the Einstein-Hilbert action written by the scalar curvature $R$ 
is promoted to an appropriate function of $R$. 
Accordingly, 
the action of $F(T)$ gravity is represented as~\cite{Linder:2010py}
$I= \int d^4x |e| \left[ F(T)/\left(2{\kappa}^2\right) 
+{\mathcal{L}}_{\mathrm{M}} \right]$ 
with $|e| = \det \left(e^A_\mu \right)=\sqrt{-g}$ and 
${\mathcal{L}}_{\mathrm{M}}$ the matter Lagrangian. 
If $F(T) = T$, this action is equivalent to that 
for pure teleparallelism. 

We assume the flat Friedmann-Lema\^{i}tre-Robertson-Walker (FLRW) 
space-time. The metric is given by 
$ds^2 = dt^2 - a^2(t) \sum_{i=1,2,3}\left(dx^i\right)^2$. 
Here, $a(t)$ is the scale factor, and 
the Hubble parameter reads $H = \dot{a}/a$, 
where the dot means the time derivative. 
In the FLRW background, we obtain the expressions of the metric 
$g_{\mu \nu}= \mathrm{diag} (1, -a^2, -a^2, -a^2)$, 
the tetrad components $e^A_\mu = (1,a,a,a)$, 
and the torsion scalar $T=-6H^2$. 

Moreover, in this background, 
the gravitational field equations are written as 
$H^2 = \left({\kappa}^2/3\right) 
\left(\rho_{\mathrm{M}}+\rho_{\mathrm{DE}} \right)
$ 
and
$\dot{H} = -\left({\kappa}^2/2\right) 
\left(\rho_{\mathrm{M}} + P_{\mathrm{M}} + 
\rho_{\mathrm{DE}} + P_{\mathrm{DE}} \right)
$. Here, $\rho_{\mathrm{M}}$ and 
$P_{\mathrm{M}}$ are the energy density and pressure 
for all of the matters, i.e., the perfect fluids, respectively. 
The continuity equation for the perfect fluid becomes 
$\dot{\rho}_{\mathrm{M}}+3H\left( \rho_{\mathrm{M}} + P_{\mathrm{M}} \right)
=0$.
Furthermore, $\rho_{\mathrm{DE}}$ and 
$P_{\mathrm{DE}}$ are the energy density and pressure 
for the dark energy components, respectively, given by 
$\rho_{\mathrm{DE}} = 
\left[1/\left(2{\kappa}^2\right) \right] J_1$ 
and 
$P_{\mathrm{DE}} =  
-\left[1/\left(2{\kappa}^2\right) \right] 
\left( 4J_2 + J_1 \right)
$ with 
$J_1 \equiv -T -F(T) +2TF'(T)$ and 
$J_2 \equiv \left( 1 -F'(T) -2TF''(T) \right) \dot{H}$, 
where the prime denotes the derivative with respect to 
$T$ as $F'(T) \equiv dF(T)/dT$ and $F''(T) \equiv d^2F(T)/dT^2$. 
The continuity equation for the dark energy components 
reads 
$\dot{\rho}_{\mathrm{DE}}+3H \left( 
\rho_{\mathrm{DE}} + P_{\mathrm{DE}}
\right) = 0$.

\begin{table*}[tbp]
\caption{Conditions that there exist 
the finite-time future singularities for $H$ in Eqs.~(\ref{eq:3.1}) and (\ref{eq:3.2}), those for $\rho_{\mathrm{DE}}$ and $P_{\mathrm{DE}}$, and 
the evolutions of $H$ and $\dot{H}$ for $t \to t_{\mathrm{s}}$. 
}
\begin{center}
\begin{tabular}
{lllll}
\hline
$q (\neq 0, \, -1)$ [Type of singularities]
& $H$ ($t \to t_{\mathrm{s}}$) \quad 
& $\dot{H}$ ($t \to t_{\mathrm{s}}$)
& $\rho_{\mathrm{DE}}$
& $P_{\mathrm{DE}}$
\\[0mm]
\hline
$q \geq 1$ [Type I (``Big Rip'')] 
& $H \to \infty$
& $\dot{H} \to \infty$
& $J_1 \neq 0$ \quad 
& $J_1 \neq 0$ or $J_2 \neq 0$  
\\[0mm]
$0 < q < 1$ [Type III] 
& $H \to \infty$
& $\dot{H} \to \infty$
& $J_1 \neq 0$ \quad 
& $J_1 \neq 0$
\\[0mm]
$-1 < q < 0$ [Type II (``Sudden'')] \quad 
& $H \to H_{\mathrm{s}}$
& $\dot{H} \to \infty$
& 
& $J_2 \neq 0$
\\[0mm]
$q < -1$ ($q$ is non-integer) [Type IV] 
& $H \to H_{\mathrm{s}}$
& $\dot{H} \to 0$
& 
& 
\\[0mm]
& 
& Divergence of higher derivatives of $H$
&
&
\\
\hline 
\end{tabular}
\end{center}
\label{tb:Table1}
\end{table*}

\subsection{Crossing of the phantom divide} 

As an $F(T)$ gravity model in which 
the crossing of the phantom divide can occur, we obtain
\begin{eqnarray} 
F(T) \Eqn{=} T + \gamma \left\{ T_0 \left( \frac{uT_0}{T}\right)^{-1/2} \ln 
\left(\frac{uT_0}{T}\right) 
-T \left[1-\exp\left(\frac{uT_0}{T}\right) \right]
\right\}\,,
\label{eq:2.1} \\ 
\gamma \Eqn{\equiv} \frac{1-\Omega_{\mathrm{m}}^{(0)}}{ 
2u^{-1/2}+\left[1-\left(1-2u\right) \exp\left(u\right) \right]}\,,
\label{eq:2.2} 
\end{eqnarray}
with $T_0$ the present value of the torsion scalar $T$ 
and $u$ a constant. 
In addition, 
$\Omega_{\mathrm{m}}^{(0)} \equiv 
\rho_{\mathrm{m}}^{(0)}/\rho_{\mathrm{crit}}^{(0)}$. 
Here, 
$\rho_{\mathrm{m}}^{(0)}$
is the current energy density of non-relativistic matter, 
and 
$\rho_{\mathrm{crit}}^{(0)} = 3H_0^2/\kappa^2$ is the current critical 
density, where $H_0$ is the Hubble parameter at the present time. 
The model in Eq.~({\ref{eq:2.1}}) consists of both the logarithmic and exponential terms. 
The EoS for dark energy is defined as $w_{\mathrm{DE}} \equiv P_{\mathrm{DE}} / \rho_{\mathrm{DE}}$. 
It is seen that for the model in Eq.~({\ref{eq:2.1}}), 
the EoS for dark energy $w_{\mathrm{DE}}$ evolves 
from $w_{\mathrm{DE}} > -1$ to $w_{\mathrm{DE}} < -1$, 
and thus the crossing of the phantom divide line $w_{\mathrm{DE}}=-1$ 
can happen. 
We remark that this manner is opposite to the representative behavior 
for $F(R)$ gravity models. 

Furthermore, it can numerically be demonstrated that for the model in Eq.~({\ref{eq:2.1}}), first the density parameter of radiation $\Omega_\mathrm{r} 
\equiv \rho_\mathrm{r} / \rho_{\mathrm{crit}}$ dominates, 
and then the density parameter of non-relativistic matter 
$\Omega_\mathrm{m} 
\equiv \rho_\mathrm{m} / \rho_{\mathrm{crit}}$ is dominant, 
and eventually the density parameter of dark energy $\Omega_\mathrm{DE} 
\equiv \rho_\mathrm{DE} / \rho_{\mathrm{crit}}$ 
becomes much larger than the density parameters of radiation and 
non-relativistic matter around the present time. 
Here, $\rho_\mathrm{r}$, $\rho_\mathrm{m}$, $\rho_\mathrm{DE}$, 
and $\rho_{\mathrm{crit}} \equiv 3H^2/\kappa^2$ 
are the energy density of radiation, that of non-relativistic matter, 
that of dark energy, and the critical density, respectively. 
Hence, in the model in Eq.~({\ref{eq:2.1}}), the dark energy dominated 
stage, which follows the radiation dominated stage and 
the matter dominated stage, can be realized. 
Moreover, through the statistical analysis with 
the recent cosmological observational data in terms of 
SNe Ia, BAO, and the CMB radiation, we derive the observational 
constraints on the model parameters of the model in Eq.~({\ref{eq:2.1}}). 
As a result, we find that the model in Eq.~({\ref{eq:2.1}}) can fit 
the observational data well. 
In Ref.~\cite{Wu:2010av}, other $F(T)$ gravity models 
in which the crossing of the phantom divide can occur 
have been built up.

\section{Finite-time future singularities}

In this section, 
we show that the finite-time future singularities can occur in $F(T)$ gravity 
by reviewing the consequences in Ref.~\cite{Bamba:2012vg}, 
We also reconstruct $F(T)$ gravity models in which the finite-time future singularities appear. 

\subsection{Four types of the finite-time future singularities} 

For the FLRW space-time, 
the effective EoS 
is given by~\cite{Nojiri:2010wj,Nojiri:2006ri} 
$ 
w_{\mathrm{eff}} \equiv P_{\mathrm{eff}}/\rho_{\mathrm{eff}} = 
-1 - 2\dot{H}/\left(3H^2\right)
$, where 
$
\rho_{\mathrm{eff}} \equiv 3H^2/\kappa^2
$ and 
$
P_{\mathrm{eff}} \equiv -\left(2\dot{H}+3H^2\right)/\kappa^2
$ 
are the energy density and pressure of all of the energy components 
in the universe, respectively. 
When the dark energy density becomes dominant over the energy density of non-relativistic matters, the following approximation is satisfied: 
$w_{\mathrm{DE}} \approx w_{\mathrm{eff}}$. 
In what follows, we explore such a situation in order to 
examine the cosmic evolution when there appear 
the finite-time future singularities at $t = t_\mathrm{s}$. 
If $\dot{H} < 0\ (>0)$, 
the universe is in the non-phantom [i.e., quintessence] (phantom) phase 
with $w_\mathrm{eff} >-1\ (<-1)$. 
For $w_\mathrm{eff} =-1$, we have $\dot{H} = 0$, namely, 
the cosmological constant. 

The finite-time future singularities are classified into 
four types~\cite{Nojiri:2005sx}. 
Type I:\ 
When $t\to t_{\mathrm{s}}$, 
$a \to \infty$,
$\rho_{\mathrm{eff}} \to \infty$ and
$\left|P_{\mathrm{eff}}\right| \to \infty$. 
This types includes the case that 
$\rho_\mathrm{{eff}}$ and $P_{\mathrm{eff}}$ become finite values 
at $t = t_{\mathrm{s}}$~\cite{Shtanov:2002ek}. 
Type II:\ 
When $t\to t_{\mathrm{s}}$, 
$a \to a_{\mathrm{s}}$, 
$\rho_{\mathrm{eff}} \to \rho_{\mathrm{s}}$ and 
$\left|P_{\mathrm{eff}}\right| \to \infty$, 
where 
$a_{\mathrm{s}} (\neq 0)$ and $\rho_{\mathrm{s}}$ are constants. 
Type III:\ 
When $t\to t_{\mathrm{s}}$, 
$a \to a_{\mathrm{s}}$, 
$\rho_{\mathrm{eff}} \to \infty$ and
$\left|P_{\mathrm{eff}}\right| \to \infty$. 
Type IV:\ 
When $t\to t_{\mathrm{s}}$, 
$a \to a_{\mathrm{s}}$, 
$\rho_{\mathrm{eff}} \to 0$, 
$\left|P_{\mathrm{eff}}\right| \to 0$. 
Here, the higher derivatives of $H$ also diverge, 
and the case that $\rho_{\mathrm{eff}}$ and/or $\left|P_{\mathrm{eff}}\right|$ 
approach finite values in the limit $t \to t_{\mathrm{s}}$ is included. 
Type I and Type II are known as ``Big Rip''~\cite{Caldwell:2003vq,McInnes:2001zw} and ``Sudden''~\cite{Barrow:2004xh,Nojiri:2004ip} singularities.

\subsection{Conditions for the finite-time future singularities to appear} 

We suppose that $H$ is described as~\cite{Nojiri:2008fk} 
\begin{eqnarray}
H \Eqn{\sim} \frac{h_{\mathrm{s}}}{\left( t_{\mathrm{s}} - t 
\right)^{q}} \,\,\, \mathrm{for} \,\,\, q > 0\,, 
\label{eq:3.1} \\
H \Eqn{\sim} H_{\mathrm{s}} + \frac{h_{\mathrm{s}}}{ 
\left( t_{\mathrm{s}} - t \right)^{q}} \,\,\, \mathrm{for} \,\,\, q<-1, 
\,\,\, -1< q < 0\,, 
\label{eq:3.2}
\end{eqnarray}
with $h_{\mathrm{s}} (> 0)$, $H_{\mathrm{s}} (> 0)$, and $q (\neq 0, \, -1)$ 
constants. 
Since the value of $H$ has to be a real number, we examine the 
range $0< t < t_{\mathrm{s}}$. 
In Table~\ref{tb:Table1}, 
we summarize the conditions that there exist 
the finite-time future singularities for $H$ in Eqs.~(\ref{eq:3.1}) and (\ref{eq:3.2}), those for $\rho_{\mathrm{DE}}$ and $P_{\mathrm{DE}}$, and 
the evolutions of $H$ and $\dot{H}$ for $t \to t_{\mathrm{s}}$.

\begin{table*}[tbp]
\caption{
Conditions that the parameters in a power-law expression for 
$F(T)$, for which the finite-time 
future singularities exist, 
and the forms of a power-low the correction term 
$F_{\mathrm{c}} (T) = B T^\beta$ curing the finite-time future singularities.}
\begin{center}
\begin{tabular}
{lcll}
\hline 
$q (\neq 0, \, -1)$ [Type of singularities]
& Consequence
& $F(T) = A T^\alpha$ ($A \neq 0$, $\alpha \neq 0$)
& $F_{\mathrm{c}} (T) = B T^\beta$ ($B \neq 0$, $\beta \neq 0$) 
\\[0mm]
\hline
$q \geq 1$ [Type I (``Big Rip'')] 
& appears
& $\alpha < 0$
& $\beta > 1$
\\[0mm]
$0 < q < 1$ [Type III] 
& ---
& $\alpha < 0$
& $\beta > 1$
\\[0mm]
$-1 < q < 0$ [Type II (``Sudden'')]
& ---
& $\alpha = 1/2$
& $\beta \neq 1/2$
\\[0mm]
$q < -1$ ($q$ is non-integer) [Type IV] 
\,\,\,\,\,
& appears
& $\alpha = 1/2$
& $\beta \neq 1/2$
\\
\hline 
\end{tabular} 
\end{center}
\label{tb:Table2}
\end{table*}

\begin{table*}[tbp]
\caption{$H$ and $F(T)$ for which 
(i) inflation in the early universe, (ii) the $\Lambda$CDM model, 
(iii) Little Rip cosmology and (iv) Pseudo-Rip cosmology can be realized. 
Here, $h_{\mathrm{inf}}$, $\Lambda$, $H_{\mathrm{LR}}$, $\zeta$, and 
$H_{\mathrm{PR}}$ are constants. 
}
\begin{center}
\begin{tabular}
{lll}
\hline 
Scenario
& $H$ 
& $F(T)$ 
\\[0mm]
\hline
(i) Power-law inflation (when $t \to 0$)
& $H = h_{\mathrm{inf}}/t$\,, $h_{\mathrm{inf}} (> 1)$
& $F(T) = A T^\alpha$\,, $\alpha <0$ or $\alpha = 1/2$
\\[0mm]
(ii) $\Lambda$CDM model or exponential inflation
& $H =\sqrt{\Lambda/3} = \mathrm{constant}$\,, $\Lambda >0$
& $F(T) = T - 2\Lambda$\,, $\Lambda >0$
\\[0mm]
(iii) Little Rip scenario (when $t \to \infty$)
& $H = H_{\mathrm{LR}} \exp \left( \zeta t \right)$\,, $H_{\mathrm{LR}} > 0$ and $\zeta >0$ 
& $F(T) = A T^\alpha$\,, $\alpha <0$ or $\alpha = 1/2$ 
\\[0mm]
(iv) Pseudo-Rip scenario\,\,\,\,\,
& $H = H_{\mathrm{PR}} \tanh \left(t/t_0\right)$\,, \,\, $H_{\mathrm{PR}} > 0$& $F(T) = A \sqrt{T}$ 
\\
\hline 
\end{tabular}
\end{center}
\label{tb:Table3}
\end{table*}

If $H$ is represented as in Eqs.~(\ref{eq:3.1}) and (\ref{eq:3.2}), 
we reconstruct $F(T)$ gravity models, in which  
the finite-time future singularities happen, 
by using the procedure~\cite{Nojiri:2006gh,Nojiri:2006be}. 
It is seen that in the flat FLRW universe, 
both of two gravitational filed equations can be met 
when $F(T)$ is given by the following power-law expression: 
$F(T) = A T^{\alpha}$, where 
$A (\neq 0)$ and $\alpha (\neq 0)$ are constants. 
Furthermore, we find a correction term $F_{\mathrm{c}} (T)$ 
curing the finite-time future singularities, given by 
$F_{\mathrm{c}} (T) = B T^\beta$ 
with $B (\neq 0)$ and $\beta (\neq 0)$ constants. 
It is known that the finite-time future singularities 
can be removed by the quadratic term (namely, $\beta = 2$)
~\cite{Nojiri:2010wj,Nojiri:2006ri} for $F(R)$ gravity 
and non-local gravity~\cite{Bamba:2012ky}. 
As a result, for $F(T) = A T^{\alpha} + B T^\beta$, 
which is the summation of the original and correction terms, 
two gravitational filed equations cannot simultaneously be satisfied. 
It follows from this fact that 
the finite-time future singularities can be removed 
by such a power-law correction term. 
We show the conditions that the 
parameters in a power-law expression for 
$F(T)$, for which the finite-time 
future singularities exist, 
and the forms of a power-low the correction term curing 
the finite-time future singularities in Table~\ref{tb:Table2}. {}From this table, it is clearly seen that Type I and IV singularities can appear.

\subsection{Various cosmological scenarios}

Moreover, we examine which kinds of 
the finite-time future singularities occur in each cosmology. 
If the absolute value of $q$ is large enough, 
the finite-time future singularities can appear. 
We also explore $F(T)$ gravity models in which 
the following cosmological scenarios can be realized: 
(i) Power-law inflation, 
(ii) The $\Lambda$CDM model, 
(iii) 
Little Rip scenario~\cite{Frampton:2011sp,Brevik:2011mm,Frampton:2011rh}, 
and (iv) Pseudo-Rip scenario~\cite{Frampton:2011aa}. 
The expressions of $H$ and $F(T)$ in the scenarios shown above 
are presented in Table~\ref{tb:Table3}. 

In addition, we consider Little Rip scenario, 
which is a kind of a mild phantom cosmology. 
The motivation of this scenario is to remove 
the finite-time future singularities including a Big Rip singularity. 
In this scenario, the dark energy density grows as the universe evolves, 
whereas the EoS for dark energy $w_{\mathrm{DE}}$ 
becomes close to $w_{\mathrm{DE}} = -1$ from $w_{\mathrm{DE}} < -1$. 
The special feature of this scenario is that at some future time, 
bound structures are dissolved because an inertial force 
operating objects becomes large. Such a phenomenon is the so-called 
Little Rip. 

As another related cosmology, we study Pseudo-Rip scenario. 
With the Hubble parameter, cosmological scenarios can be classified~\cite{Frampton:2011aa}. 
(i) Power-law inflation: 
$H(t) \to \infty$ for $t \to 0$. 
(ii) The $\Lambda$CDM model (or Exponential inflation): 
$H(t) = H(t_0) = \mathrm{constant}$, where $t_0$ is the current time. 
(iii) Little Rip scenario: 
$H(t) \to\infty$ for $t \to \infty$. 
(iv) Pseudo-Rip scenario 
(a phantom cosmology approaching de Sitter expansion asymptotically: 
$H(t) \to H_\infty < \infty$ for $t \to \infty$ with 
$t \geq t_0$ and $H_\infty (>0)$ a constant. 
For a Big Rip singularity, 
$H(t) \to \infty$, \, $t \to t_{\mathrm{s}}$, 
as depicted in Table~\ref{tb:Table1}. 

We note that 
the EoS parameter $w_{\mathrm{DE}}$ for dark energy, 
the deceleration parameter $q_{\mathrm{dec}} 
\equiv -\ddot{a}/\left(aH^2\right)$,  
the jerk parameter $j \equiv \dddot{a}/\left(aH^3\right)$ and 
the snark parameter $s \equiv \left(j - 1\right)/\left[3 \left( q_{\mathrm{dec}} -1/2 \right)\right]$~\cite{Chiba:1998tc,Sahni:2002fz} 
are used to observationally constrain the dark energy models. 
For the $\Lambda$CDM model, we have 
$(w_{\mathrm{DE}}, q_{\mathrm{dec}}, j, s) = 
(-1,-1,1,0)$. 
In the flat universe, there have been proposed 
$w_{\mathrm{DE}} = -1.10 \pm 0.14 \, 
(68 \% \, \mathrm{CL})$~\cite{Komatsu:2010fb}. 
By using these parameters, especially, 
the observational constraints on 
the models parameters can be derived. 
For example, with the observational value of $w_{\mathrm{DE}}$, 
the constraints on 
$H_{\mathrm{LR}}$ and $H_{\mathrm{PR}}$ shown in Table~\ref{tb:Table3} 
can be derived. 
In Little Rip scenario, 
we obtain $H_{\mathrm{LR}} \geq \left[2H_0/\left(3e\right)\right]
\left(1/0.24\right) = 1.50 \times 10^{-42} \, \mathrm{GeV}$, 
where $H_0 = 2.1 h \times 10^{-42} \, \mathrm{GeV}$~\cite{Kolb and Turner} 
with $h = 0.7$~\cite{Komatsu:2010fb,Freedman:2000cf} is 
the current Hubble parameter and $e = 2.71828$, and 
$\chi \equiv H_0/\left(H_{\mathrm{LR}} e \right) \leq 0.36$. 
On the other hand, for Pseudo Rip scenario, 
we get $H_{\mathrm{PR}} \geq \left(2H_0/3\right) 
\left[4/\left(e - e^{-1}\right)^2\right] \left(1/0.24\right) = 2.96 
\times 10^{-42} \, \mathrm{GeV}$ 
and $\delta \equiv H_0/H_{\mathrm{PR}} \leq 0.497196$. 
Hence, Little Rip scenario with $\chi \ll 1$ 
and Pseudo Rip scenario with an appropriate value of $\delta$ 
can be compatible with the $\Lambda$CDM model.

\subsection{Inertial force}

In the expanding universe, the relative acceleration between two points 
separated by a distance $l$ is given by 
$l \ddot{a}/a$, where $a$ is the scale factor. 
Suppose that there exists a particle with mass $m$ at each of the points, 
an observer at one of the masses would measure an inertial force on the 
other mass. 
We assume that there are two particles (A) and (B) with its mass $m$ and 
the distance between them is $l$. 
The inertial force $F_{\mathrm{inert}}$ on the particle (B), 
which is measured by an observer at the point of the particle (A), 
is represented as 
$F_{\mathrm{inert}} = ml\ddot{a}/a = ml \left( \dot{H} + H^2 \right)$
~\cite{Frampton:2011sp,Frampton:2011rh}. 
In the case of a Big Rip singularity with $H$ in Eq.~(\ref{eq:3.1}), 
$F_{\mathrm{inert}} \to \infty$ when $t \to t_{\mathrm{s}}$. 
Moreover. 
for Little Rip scenario with $H$ described in Table~\ref{tb:Table3}, 
$F_{\mathrm{inert}} \to \infty$ when $t \to \infty$. 
Furthermore, 
in Pseudo-Rip scenario with $H$ presented in Table~\ref{tb:Table3}, 
$F_{\mathrm{inert}} \to F_{\mathrm{inert}\,,\infty}^{\mathrm{PR}} 
\equiv ml H_{\mathrm{PL}}^2 < \infty$ 
when $t \to \infty$. 
Thus, $F_{\mathrm{inert}}$ approaches a finite value asymptotically. 
This is because $H \to H_{\mathrm{PR}}$ and $\dot{H} \to 0$.  

If a force $F_{\mathrm{b}}$ to bound two particles is smaller than 
a positive inertial force $F_{\mathrm{inert}} (>0)$, the two particle 
bound system is disintegrated. 
As an example, we examine the Earth-Sun (ES) system. 
When $F_{\mathrm{inert}\,,\infty}^{\mathrm{PR}} > 
F_{\mathrm{b}}^{\mathrm{ES}} = G M_{\oplus} M_{\odot} / 
r_{\oplus-\odot}^2 = 4.37 \times 10^{16} \, 
\mathrm{GeV}^2$, which is the bound force in the ES system, 
the ES system is dissociated, so that Pseudo-Rip scenario can be satisfied. 
Here, we have used the mass of Earth $M_{\oplus} = 3.357 \times 10^{51} \, 
\mathrm{GeV}$~\cite{Kolb and Turner} and that of Sun $M_{\odot} = 1.116 \times 10^{57} \, \mathrm{GeV}$~\cite{Kolb and Turner}, 
and we have set $m = M_{\oplus}$ and 
$l = r_{\oplus-\odot} = 1 \mathrm{AU} = 7.5812 \times 10^{26} \, \mathrm{GeV}^{-1}$~\cite{Kolb and Turner} (the Astronomical unit, namely, the 
distance between Earth and Sun). 
In this case, we acquire $H_{\mathrm{PR}} > 
\sqrt{G M_{\odot} / r_{\oplus-\odot}^3} = 1.31 \times 10^{-31}\, 
\mathrm{GeV}$. 
This is consistent with the observations on the present value 
of $w_{\mathrm{DE}}$ in Pseudo-Rip scenario because 
this is much stronger than the constraint 
$H_{\mathrm{PR}} \geq 2.96 \times 10^{-42} \, \mathrm{GeV}$ 
given above, which originates from the observations on 
the value of $w_{\mathrm{DE}}$ at the present time.  

It is also remarked that in the process of collapse of the star, 
the time-dependent matter instability can happen 
not only for $F(R)$ gravity~\cite{Arbuzova:2010iu,Bamba:2011sm} but also $F(T)$ gravity.

\section{Higher-dimensional theories}

In this section, 
we construct four-dimensional effective $F(T)$ gravity theories from the five-dimensional Kaluza-Klein (KK)~\cite{F-M,MKK-A-C-F,Overduin:1998pn} and Randall-Sundrum (RS)~\cite{Randall:1999ee,Randall:1999vf} theories 
by following the investigations in Ref.~\cite{Bamba:2013fta}.

\subsection{Five-dimensional KK theory}

First, we derive the effective $F(T)$ gravity theories in the four-dimensional space-time from the KK theory in the five-dimensional space-time. 
It is supposed that in $F(T)$ gravity, 
the ordinary procedure of 
the KK reduction~\cite{F-M,MKK-A-C-F,Overduin:1998pn} 
can be executed from the five-dimensional space-time to the four-dimensional 
space-time. 
In this process, one dimension of space is compacted into a small circle, 
while the four-dimensional space-time is infinitely extended. 
The radius of the fifth dimension is set to be around the Planck length 
so that the KK effects cannot appear. 
Consequently, the size of the circle is small enough for the phenomena 
in the quite low energy scale not to be seen. {}From now on, 
we concentrate on the gravity sector in the action, 
and therefore the matter sector is neglected. 

The five-dimensional action in $F(T)$ gravity is~\cite{Capozziello:2012zj} 
\begin{eqnarray} 
{}^{(5)}S \Eqn{=} \int d^5 x \left|{}^{(5)}e\right| 
\frac{F({}^{(5)}T)}{2 \kappa_5^2}\,, 
\label{eq:4.1} \\ 
{}^{(5)}T \Eqn{\equiv} 
\frac{1}{4} T^{a b c}T_{a b c} + \frac{1}{2} 
T^{a b c}T_{c b a}-T_{a b}^{\verb|  |a}
T^{c b}_{\verb|  |c}\,. 
\label{eq:4.2}
\end{eqnarray}
Here, ${}^{(5)}T$ is the torsion scalar, 
the Latin indices are $(a, b, \dots = 0, 1, 2, 3, 5)$ 
with ``$5$'' the fifth-coordinate component, 
${}^{(5)}e = \sqrt{{}^{(5)}g}$ with ${}^{(5)}g$ 
the determinant of ${}^{(5)}g_{\mu\nu}$, and 
$\kappa_{5}^2 \equiv 8 \pi G_5 = 
\left( {}^{(5)}M_{\mathrm{Pl}} \right)^{-3}$ 
with $G_5$ the gravitational constant and 
$M_{\mathrm{Pl}}^{(5)}$ the Planck mass. 
The superscription ``$(5)$'' or the subscription``$5$'' 
shows the quantities in the five-dimensional space-time. 
The representation of ${}^{(5)}T$ is equivalent to 
that of the torsion scalar 
in the four-dimensional space-time. 
The five-dimensional metric is given by 
${}^{(5)}g_{\mu\nu} = \mathrm{diag} (g_{\mu\nu}, -\psi^2)$, 
where $\psi$ is a dimensionless and 
(spatially) homogeneous scalar field 
(namely, it only has the time dependence). 

The four-dimensional effective action becomes 
\begin{equation} 
S_{\mathrm{KK}}^{(\mathrm{eff})} = \int d^4x |e| 
\frac{1}{2\kappa^2} \psi 
F(T +\psi^{-2} \partial_{\mu} \psi \partial^{\mu} \psi)\,, 
\label{eq:4.3}
\end{equation}
where we have used  $e^A_a = \mathrm{diag} (1, 1, 1, 1, \psi)$ and 
$\eta_{a b} = \mathrm{diag} (1, -1, -1, -1, -1)$. 
In the case that $F(T) = T-2\Lambda_4$ with $\Lambda_4 (>0)$ 
the positive cosmological constant in 
the four-dimensional space-time, 
by defining another scalar field $\xi$ as 
$\psi \equiv \left(1/4\right) \xi^2$, 
we find that the action in Eq.~(\ref{eq:4.3}) 
is described as~\cite{F-M}
\begin{equation} 
S_{\mathrm{KK}}^{(\mathrm{eff})} |_{F(T)=T-2\Lambda_4} 
= \int d^4x |e| 
\frac{1}{\kappa^2}
\left[ \frac{1}{8} \xi^2 T + \frac{1}{2} \partial_{\mu} \xi \partial^{\mu} \xi - \Lambda_4 \right]\,.
\label{eq:4.4}
\end{equation}
In the flat FLRW background, from the action in Eq.~(\ref{eq:4.4}), 
the gravitational field equations are given by 
$\left(1/2\right) \dot{\xi}^2 
-\left(3/4\right) H^2 \xi^2 + \Lambda_4 = 0$ and 
$\dot{\xi}^2 + H \xi \dot{\xi} 
+\left(1/2\right) \dot{H} \xi^2 = 0$~\cite{Geng:2011aj}, 
and the equation of motion in terms of $\xi$ is written as 
$\ddot{\xi} + 3H\dot{\xi} + \left(3/2\right) H^2 \xi = 0$. 
By using the gravitational field equations, we have 
$\left(3/2\right) H^2 \xi^2 -2\Lambda_4 + H \xi \dot{\xi} 
+ \left(1/2\right) \dot{H} \xi^2 = 0$. 
Its solution is 
$H = H_{\mathrm{inf}} = \mathrm{constant} (>0)$, 
where $H_{\mathrm{inf}}$ is 
the Hubble parameter at the inflationary stage, 
and $\xi = \xi_1 \left(t/\bar{t}\right) + \xi_2$ 
with $\xi_1$ and $\xi_2 (>0)$ constants, 
where $\bar{t}$ denotes a time. 
Thus, when $t \to 0$, 
inflation with the de Sitter expansion can be realized, where 
$H_{\mathrm{inf}} \approx \left(2/\xi_2\right) \sqrt{\Lambda_4/3}$, 
$a \approx \bar{a} \exp \left( H_{\mathrm{inf}} t \right)$,  
and $\xi \approx \xi_2$. 
Moreover, with the equation of motion in terms of $\xi$, 
we acquire $\xi_1 \approx -\left(1/2\right) \xi_2 H_{\mathrm{inf}} \bar{t} 
\approx -\sqrt{\Lambda_4/3} \bar{t}$.

\subsection{RS brane-world model}

Next, we deduce the effective $F(T)$ gravity theories in the 
four-dimensional space-time from the RS brane-world model in the five-dimensional space-time. 
There exist two branes in the RS type-I model~\cite{Randall:1999ee}: 
A positive tension brane located at $y=0$ and a negative one located at $y=u$, 
where $y$ means the fifth dimension. 

The five-dimensional metric is expressed as 
\begin{equation} 
ds^2 = \exp \left(-2\frac{|y|}{l}\right) g_{\mu\nu} (x) 
dx^{\mu} dx^{\nu} + dy^2 \,,
\label{eq:4.5}
\end{equation}
where $l=\sqrt{-6/\Lambda_5}$, 
$\exp \left( -2|y|/l \right)$ is the warp factor, and $\Lambda_5 (< 0)$ is 
the negative cosmological constant in the bulk. 
When $u \to \infty$, 
the RS type-I model is reduced to 
the RS type-II model~\cite{Randall:1999vf}. 
In this model, 
there is only one positive tension brane 
in the anti-de Sitter bulk space. 
In Ref.~\cite{Shiromizu:1999wj}, 
the gravitational field equation on the brane 
has been presented for the RS type-II model. 
It corresponds to the induced equation, i.e., the Gauss-Codazzi equation, on the brane, and it is derived by using 
the Israel's junction conditions on the brane and the $Z_2$ symmetry of $y \leftrightarrow -y$. 
This procedure has recently been considered in teleparallelism~\cite{Nozari:2012qi}. 
The vector part of the torsion tensor in the bulk is projected on the brane, 
so that new terms, which do not exist in the curvature gravity, can emerge. 

In the flat FLRW background, 
the Friedmann equation on the brane is given by 
\begin{equation} 
H^2 \frac{d F(T)}{dT} = -\frac{1}{12} 
\left[ F(T) - 4 \Lambda 
-2 \kappa^2 \rho_{\mathrm{M}} 
- \left(\frac{\kappa_5^2}{2}\right)^2 \mathcal{I} \rho_{\mathrm{M}}^2 
\right]\,.
\label{eq:4.6}
\end{equation}
Here, 
$\mathcal{I} \equiv \left(11-60 w_{\mathrm{M}} +93 w_{\mathrm{M}}^2 \right)/4$ with 
$w_{\mathrm{M}} \equiv P_{\mathrm{M}}/\rho_{\mathrm{M}}$ 
the EoS for matter, where $\rho_{\mathrm{M}}$ and $P_{\mathrm{M}}$ are 
the energy density and pressure of matter, respectively. 
In the expression of $\mathcal{I}$, 
the novel contributions in teleparallelism 
are included (there are not these terms in the curvature gravity). 
The effective cosmological constant on the brane reads 
$\Lambda \equiv \Lambda_5 + \left(\kappa_5^2/2\right)^2 \lambda^2$, 
where $\lambda (> 0)$ is the brane tension 
and we obtain the relation $G = \left[1/\left(3\pi \right) \right] \left(\kappa_5^2/2\right)^2 \lambda$. 

In the following, we consider the situation that 
the dark energy is dominant and hence 
the contribution of matter is negligible. 
If $F(T) = T - 2 \Lambda_5$, 
with $T=-6H^2$, 
we get a solution of the de Sitter expansion as 
$H = H_{\mathrm{DE}} = 
\sqrt{\Lambda_5+\kappa_5^4\lambda^2/6}$ 
and $a(t) = a_{\mathrm{DE}} \exp \left( H_{\mathrm{DE}} t \right)$, 
where $H_{\mathrm{DE}}$ and 
$a_{\mathrm{DE}} (>0)$ are constants. 
Accordingly, the late-time cosmic acceleration can happen. 
In addition, 
when $F(T) = \left(T^2/\bar{M}^2\right) + \eta \Lambda_5$ 
with $\bar{M}$ a mass scale and $\eta$ a constant, 
we find a de Sitter solution with the constant Hubble parameter
\begin{equation} 
H = H_{\mathrm{DE}} = 
\left\{ \frac{\bar{M}^2}{108} 
\left[ \left(\eta-4\right) \Lambda_5 
-4 \left(\frac{\kappa_5^2}{2}\right)^2 \lambda^2 
\right] \right\}^{1/4}\,.
\label{eq:4.7}
\end{equation}
In this expression, 
the content of the 4th root has to be positive. 
Therefore, we obtain a constraint on $\eta$, given by 
$\eta \geq 4 + \left(\kappa_5^2 \lambda^2\right)/\Lambda_5$.

\section{Summary}

In the present paper, we have stated various cosmological issues as well as theoretical properties in $F(T)$ gravity. 
First, we have investigated the cosmological evolution of the EoS for dark energy $w_{\mathrm{DE}}$. We have constructed an $F(T)$ gravity 
model consisting of an exponential term and a logarithmic one, 
in which the crossing of the phantom divide can occur.  

Next, we have found that the Type I and IV finite-time future singularities 
can appear, and reconstructed an $F(T)$ gravity model in which there exist
the finite-time future singularities. 
We have also demonstrated that by adding a power-law term of 
$T^{\beta} \, (\beta >1)$ like $T^2$,  
the finite-time future singularities can be cured, 
similarly to that in $F(R)$ gravity. 
Furthermore, we have explored $F(T)$ gravity models in which the following cosmological scenario is satisfied: power-law inflation, the $\Lambda$CDM model, 
the Little Rip scenario, and the Pseudo Rip scenario.  

Moreover, we have analyzed 
four-dimensional effective action of $F(T)$ gravity 
originating from the five-dimensional KK theories and RS models. 
We have derived the four-dimensional effective action with 
a coupling of the torsion scalar to a scalar field through 
the KK reduction to the four-dimensional space-time from 
the five-dimensional one. We have shown that in this theory, 
inflation can occur. 
We have also found that in the RS type-II model with the four-dimensional FLRW brane, $F(T)$ gravity influences on the four-dimensional FLRW brane. 
We have seen that for this model, the late-time cosmic acceleration 
can be realized. Here, inflation or the late-time cosmic acceleration can happen thanks to the torsion effect, and not by the curvature one, so that these KK theories and RS models can be considered to be constructed by not 
the scalar curvature but the torsion one in teleparallel gravity.   

It should be cautioned that there is no local Lorentz invariance in $F(T)$ gravity as indicated in Refs.~\cite{Li:2010cg,Sotiriou:2010mv}, 
and this theory is acausal~\cite{C-I-N-O,I-G-O,O-I-N-C}. 
These are the most crucial points for $F(T)$ gravity theory. 
Thus, these problems have to further be considered seriously.  

Finally, we mention a number of other cosmological subjects 
have been studied in $F(T)$ gravity. As examples, the authors works 
are raised: Trace-anomaly driven inflation~\cite{Bamba:2014zra}, 
gravitational wave modes~\cite{I-K,Bamba:2013ooa}, conformal symmetry~\cite{Bamba:2013jqa}, thermodynamics~\cite{Bamba:2011pz,Bamba:2012rv}, and 
the generation of the large-scale magnetic 
fields~\cite{Bamba:2012mi,Bamba:2013rra}\footnote{Also in $F(R)$ gravity, 
trace-anomaly driven inflation~\cite{Bamba:2014jia,Bamba:2014mua}, thermodynamics~\cite{Bamba:2009ay,Bamba:2009gq,Bamba:2009id,Bamba:2011jq,Bamba:2010kf}, and 
the generation of the large-scale magnetic fields~\cite{Bamba:2008ja,Bamba:2008xa} have been studied.}. 
It is expected that through such various cosmological investigations 
in $F(T)$ gravity, the clues to find novel viability conditions for 
$F(T)$ gravity as an alternative gravity theory to general relativity 
can be acquired.

\section*{Acknowledgments}

The author sincerely thank the organizers including 
Professor Dr. Muhammad Sharif (Chair) and Professor Dr. Asar Qadir 
for their very kind invitation to International Conference on Relativistic Astrophysics held on 10th--14th February, 2015, 
at Department of Mathematics, University of the Punjab, Lahore, Pakistan, 
and their quite hearty hospitality very much. 
Furthermore, he is really grateful to all of the 
collaborators of our studies 
in terms of $F(T)$ gravity, especially, all the co-authors of the works in Refs.~\cite{Bamba:2010wb,Bamba:2010iw,Bamba:2012vg,Bamba:2013fta}: Professor Chao-Qiang Geng, Dr. Chung-Chi Lee, Dr. Ling-Wei Luo, Professor Ratbay Myrzakulov, Professor Shin'ichi Nojiri, and Professor Sergei D. Odintsov, the results obtained in which are the basis of this paper. 
The work of the author was partially supported by the JSPS Grant-in-Aid for Young Scientists (B) \# 25800136.



\end{document}